\documentclass[final,5p,times,twocolumn]{elsarticle}

\usepackage{amssymb,amsmath,bm,mathtools}
\usepackage{siunitx,booktabs}
\usepackage{graphics,graphicx,epstopdf,epsfig}
\usepackage{tabularx}
\usepackage{multirow}
\usepackage{psfrag}
\usepackage{array}
\usepackage{hyperref}
\usepackage[english]{babel}

\def\nbR{\ensuremath{\mathrm{I\! R}}}
\newcommand{\ra}[1]{\renewcommand{\arraystretch}{#1}}
\newcommand{\be}{\begin{equation}}
\newcommand{\ee}{\end{equation}}
\newcommand{\ba}{\begin{eqnarray}}
\newcommand{\ea}{\end{eqnarray}}

\journal{Physics Letters B}

\begin{document}

\begin{frontmatter}



\title{Connecting spatial moments and momentum densities}

\author[label1]{M.~Hoballah}  
\author[label2]{M.B.~Barbaro}
\author[label1]{R.~Kunne.     }
\author[label1]{M.~Lassaut.  }
\author[label1]{D.~Marchand} 
\author[label3]{G.~Qu\'em\'ener}
\author[label1]{E.~Voutier}
\author[label1]{J.~van~de~Wiele }
\address[label1]{Universit\'e Paris-Saclay, CNRS/IN2P3, IJCLab, 91405 Orsay, France}
\address[label2]{Dipartimento di Fisica, Universit\'a di Torino and INFN Sezione di Torino, 10125 Torino, Italy }
\address[label3]{Normandie Univ, ENSICAEN, UNICAEN, CNRS/IN2P3, LPC Caen, 14000 Caen, France}


\begin{abstract}
The precision of experimental data and analysis techniques is a key feature of any discovery attempt. A striking example is the proton radius puzzle where the accuracy of the spectroscopy of muonic atoms challenges traditional electron scattering measurements. The present work proposes a novel method for the determination of spatial moments from densities expressed in the momentum space. This method  provides a direct access to even, odd, and more generally any real, negative and positive moment with order larger than $-3$. As an illustration, the application of this method to the electric form factor of the proton is discussed in detail.  
\end{abstract}




\end{frontmatter}


\section{Introduction}

The determination of the proton charge radius $r_E$ from the proton electric form factor measured experimentally through the elastic scattering of electrons off protons is the subject of an intense scientific activity (see Ref.~\cite{Car15,Hil17} for recent reviews). According to the definition 
\begin{equation} r_E \equiv \sqrt{ -6 \left. \frac{\mathrm{d} G_E(k^2)}{\mathrm{d}k^2} \right\vert_{k^2=0} } \, , \label{redef} \end{equation}
the experimental method to determine $r_E$ in subatomic physics consists in the evaluation of the derivative of the electric form factor of the proton $G_E(k^2)$ at zero-momen\-tum transfer. Consequently, the method strongly relies on the zero-momentum extrapolation of the $k^2$-dependency of the electric form factor measured in elastic lepton scattering off protons. In light of the proton radius puzzle~\cite{Ber14} originating from the disagreement between electron scattering~\cite{Ber10} and muonic spectroscopy~\cite{Poh10} measurements, this method has been scrutinized in every respect to suggest that the extrapolation procedure of experimental data to zero-momentum transfer suffers from  limited accuracy. The derivative method is very sensitive to the functional used to perform the extrapolation and to the upper limit of the $k^2$ momentum domain considered for this purpose~\cite{Lee15}. 
%The significant difference between the  proton charge radii, obtained on the one hand from electron elastic scattering (0.879(8)~fm~\cite{Ber10}), and on the other hand from the spectroscopy of muonic hydrogen (0.84184(67)~fm~\cite{Poh10}), is induced by a very small difference in the electric form factor values at very low momentum transfers. This puts unbearable constraints on the systematics of lepton scattering experiments~\cite{Sic17}. 
The significant difference between the proton charge radius obtained from electron elastic scattering (0.879(8)~fm~\cite{Ber10}) and that obtained from the spectroscopy of muonic hydrogen (0.84184(67)~fm~\cite{Poh10}) implies such a small difference in the electric form factor values at very low momentum transfers that it puts unbearable constraints on the systematics of lepton scattering experiments~\cite{Sic17}.
As a matter of fact, the precision of the highest quality electron scattering  measurements~(0.879(8)~fm~\cite{Ber10} and 0.831(14)~fm~\cite{Xio19}) on that issue remains $\sim$10 times worse than that of muonic atom  measurements~\cite{Ant13,Poh16}. Improving the precision of the so-called derivative method to such a competitive level does not appear reachable with current knowledge and technologies~\cite{Hob19}.

Within a non-relativistic description of the internal structure of the proton (see Ref.~\cite{Mil19} for a recent discussion of relativistic effects), Eq.~\ref{redef} can be recovered from the MacLaurin expansion of the electric form factor expressed as the Fourier transform of the proton charge density $\rho_E({\bold r})$,
\begin{equation}
G_E(k^2) = \int_{\nbR^3} \mathrm{d}^3\bold{r} \, e^{- i \bold{k}\cdot \bold{r}} \rho_E({\bold r}) \, ,
\end{equation}
namely
\begin{equation} G_E(k^2) = \sum_{j=0}^{\infty} (-1)^j \, \frac{k^{2j}}{(2j+1)!} \, \langle r^{2j} \rangle \end{equation}
where $k$ is the Euclidian norm of $\bold{k}$. Here 
\begin{equation} \langle r^{2j} \rangle = (-1)^j \, \frac{(2j+1)!}{j!} \, \left. \frac{\mathrm{d}^j G_E(k^2)}{\mathrm{d} (k^2)^j} \right\vert_{k^2=0} \label{eq:there} \end{equation}
relates the electric form factor to the even moments $\langle r^{2j} \rangle$ of the charge density $\rho_E({\bold r})$
\be
\langle r^{2j} \rangle \equiv (r^{2j},\rho_E) = \int_{\nbR^3}  \mathrm{d}^3\bold{r} \, r^{2j} \rho_E({\bold r})\, .
\ee
Consequently, the non-relativistic charge radius of the proton may be expressed as
\begin{equation} r_E = \sqrt{ \langle r^2 \rangle } \, .\end{equation}
The discrepancies between the latest scattering measurements of the proton radius~\cite{Ber10, Xio19, Mih17} clearly indicate the experimental difficulty in measuring the first derivative of the form factor. Additionally, moments of the charge density beyond the second order are also of interest as they carry complementary information on the charge distribution inside the proton. However, beyond the limited precision of the experimental determination of the $j^{\mathrm{th}}$ derivative of the form factor, the derivative method accesses only even moments of the density.

The purpose of the current work is to propose a new and intrinsically more accurate method for the determination of the spatial moments of a density from momentum space experimental observables, assuming that only the Fourier transform of the probability density function is known. This method allows access to both odd and even, positive and negative, moments of the distribution and it overcomes the limitations of the derivative technique. Its advantage lies in the more precise determination of spatial moments through integral forms of the Fourier transform of the distribution. These are expected to be less dependent on point-to-point systematics and hence more precise. The validity of this approach is demonstrated on the basis of generic densities, and its importance in the experimental determination of physics quantities is further discussed. The method for a generic probability distribution is described in Sec.~\ref{sec:spatmom}, presenting two different  regularization schemes for the Fourier transform yielding the spatial moments. The applicability of the method to a specific physical problem is discussed in Sec.~\ref{sec:meth}. The possible applications of the method to experimental data are outlined in Sec.~\ref{sec:app}, and conclusions are drawn in Sec.~\ref{sec:concl}.

%
% ------------------------------------------------------------------------------------------------------------------------
%
\section{Spatial moments}
\label{sec:spatmom}

Let $f(\bold{r})$ be a fastly decreasing function in the $3$-dimensional space. Without any loss of generality for the present discussion (see~\ref{App0}), $f(\bold{r}) \equiv f(r)$ is assumed to be a pure radial function normalized to the constant $\tilde{f}_0$ 
\begin{equation}
\int_{\nbR^3} \mathrm{d}^3\bold{r} \, f(\bold{r}) = 4 \pi \int \mathrm{d}r \, r^2 f(r) = \tilde{f}_0 \, . 
\end{equation}
Its Fourier transform 
\begin{equation}
\tilde f(\bold{k}) \equiv \tilde f(k) = \int_{\nbR^3} \mathrm{d}^3\bold{r} \, e^{- i \bold{k}\cdot \bold{r}} f(\bold{r}) 
\end{equation}
exists for any values of $k$. When $\tilde f(\bold{k})$ is integrable over $\nbR^3$, the inverse Fourier transform exists and is defined by 
\begin{equation}
f(\bold{r}) \equiv f(r) = \frac{1}{(2\pi)^3} \int_{\nbR^3} \mathrm{d}^3\bold{k} \, e^{i \bold{k}\cdot \bold{r}} \tilde f(\bold{k}) \, . \label{eq:inv} 
\end{equation}
The moments $(r^{\lambda},f)$ of the operator $r$ for the function $f$ are defined by~\cite{Gue62}
\begin{equation} \label{sect2eq01}
(r^{\lambda},f) = \int_{\nbR^3} \mathrm{d}^3\bold{r} \, r^{\lambda} \, f(\bold{r}) \, .
\end{equation}
Replacing $f({\bold r})$ with the inverse Fourier transform of $\tilde{f}({\bold k})$ (Eq.~\eqref{eq:inv}) and switching the integration order, Eq.~\ref{sect2eq01} becomes 
\begin{equation} \label{sect2eq02}
(r^{\lambda},f) = \frac{1}{(2\pi)^3}\ \int_{\nbR^3} \mathrm{d}^3\bold{k} \, \tilde{f}(\bold{k}) \int_{\nbR^3}  \mathrm{d}^3\bold{r} \, e^{i \bold{k}\cdot \bold{r}} r^{\lambda} \, .
\end{equation}
The left-hand side of Eq.~\ref{sect2eq02}, the moment $(r^{\lambda},f)$, is a finite quantity which represents a physics observable. However, the right-hand side of Eq.~\ref{sect2eq02} contains the integral  
\be \label{sect2eq03}
g_{\lambda}(\bold{k}) \equiv g_{\lambda} (k) = \int_{\nbR^3} \mathrm{d}^3\bold{r} \, e^{i \, \bold{k}\cdot \bold{r} } r^{\lambda} \, ,
\ee
that can be interpreted as the Fourier transform of the tempered distribution $r^{\lambda}$. This integral does not exist in a strict sense for $\lambda \ge -1$ but can still be treated as a distribution; the finiteness of the left-hand side ensures the physical  representativity of this expression as well as the convergence of the 6-fold integral. For instance, Eq.~\ref{sect2eq03} corresponds to the Dirac $\delta$-distribution for $\lambda$=$0$. Considering a real positive value $t$, the definition of $g_{\lambda}(\bold{k})$ provides the property 
\begin{equation}
g_{\lambda} (t\bold{k}) = \frac{1}{t^{\lambda+3}} \, g_{\lambda}(\bold{k}) \, ,  
\end{equation}
which is satisfied only by $g_{\lambda}(\bold{k})$ functions proportional to $1/k^{\lambda+3}$~\cite{Gue62,Sch65}. Eq.~\ref{sect2eq02} can then be  written as
\begin{equation}\label{sect2eq05}
(r^{\lambda},f) = \mathcal{N}_{\lambda} \int_{0} ^{\infty} \mathrm{d}k \, { \left\{ \frac{ \tilde{f}(k) } {k^{\, \lambda +1 }} \right\} } \, ,
\end{equation}
where $\mathcal{N}_{\lambda}$ is the normalization coefficient defined for $\lambda \neq 0, 2, 4...$ as
\begin{equation}
\mathcal{N}_{\lambda} = \frac{2^{\lambda+2}}{\sqrt{\pi}} \, \frac{\Gamma(\frac{\lambda+3}{2})}{\Gamma(-\frac{\lambda}{2})} 
\end{equation}
in terms of the $\Gamma$ function~\cite{Erd53}, with $\lambda>-3$. The integral in Eq.~\ref{sect2eq05} is taken in the sense of distributions, {\it i.e.} the principal value of the integral defined from the regularization of the diverging integrand at zero-momentum 
\begin{equation}\label{sect2eq6}
\left\{ \frac{ \tilde{f}(k) } {k^{\, \lambda +1 }} \right\} \equiv \frac{1}{k^{\lambda +1 }} \ \left( \tilde{f}(k) - \sum_{j=0}^{n} \tilde{f}_{2j} \, k^{2j}  \right)
\end{equation}
with 
\begin{equation} \tilde{f}_{2j} = \frac{1}{j!} \, \left. \frac{\mathrm{d}^{j} \tilde{f}(k)}{\mathrm{d}(k^2)^{j}} \right\vert_{k=0} \, . \end{equation}
Here, $n+1$ is the number of counterterms in the MacLaurin development of $\tilde{f}(k)$, where $n$=$[\lambda/2]$ is the integer part of $\lambda/2$ (with $\lambda \neq 0, 2, 4...$). It is because $\tilde{f}(k)$ originates from a pure radial function that this development is an even function of $k$. 

The right-hand side of Eq.~\ref{sect2eq05} is a convergent quantity as a whole, {\it i.e.} divergences that may appear in the normalization coefficient are compensated by the integral. The integral exists for every $\lambda$ in the domain $n < \lambda/2 < n+1$~\cite{Gue62,Sch65}, which ensures the convergence of the integrand both when $ k \to 0^+$ and when $ k \to \infty$. While the integrand diverges for even $\lambda$, even moments still accept a finite limit. Denoting for convenience $\lambda$=$m$-$\eta$ with $m$ integer, the moments $(r^{m-\eta},f)$ write 
\begin{equation} \label{sect2eq08}
(r^{m-\eta}, f ) = \mathcal{N}_{m-\eta} \  \int_{0} ^{\infty} \mathrm{d}k \ \frac{\tilde{f}(k) - \sum_{j=0}^{n} \tilde{f}_{2j} \, k^{2j}}{k^{m-\eta+1}}
\end{equation}
where $n$=$\left[ {(m-1)/2} \right]$ with $0<\eta<1$ for even values of $m$, and $0 \leq \eta <1$ for odd values of $m$. Even (odd) moments are obtained taking the limit $\eta \to 0^+$ (setting $\eta=0$). Respectively, 
\begin{eqnarray}
(r^{m},f) & = & \lim_{\eta \to 0^+} (r^{m-\eta}, f)   \qquad\quad {m \ \rm{even}} \label{sect2eq06q} \\
(r^{m},f) & = & ( r^{m-\eta},f ) \vert_{\eta=0} \, \, \qquad\quad {m \ \rm{odd}} \, .  \label{sect2eq06p}
\end{eqnarray}
The counterterms expansion of Eq.~\ref{sect2eq08} is given in Tab.~\ref{tab:counterterms} for the first order moments.     

\begin{table}[t!]
\ra{1.25}
\begin{center}
\begin{tabular}{@{}rlll@{}}
\toprule[0.95pt]
%\hline
$\pmb{m}$ & $\pmb{k^{m - \eta + 1}}$ & $\pmb{n}$ & $\pmb{\sum_{j=0}^{n} \tilde{f}_{2j} \, k^{2j}}$  \\
\midrule[0.7pt]
%\hline
-2 & $ k^{-1-\eta}$ & -2 & -                \\
-1 & $ k^{-\eta}$ & -1 &  -               \\
0 & $ k^{1-\eta}$ & -1 &  -               \\
1 & $ k^{2-\eta}$ & \phantom{-}0 & $\tilde{f}_0$        \\
2 & $ k^{3-\eta}$ & \phantom{-}0 & $\tilde{f}_0$                    \\
3 & $ k^{4-\eta}$ & \phantom{-}1 & $\tilde{f}_0+\tilde{f}_2 k^2$      \\
4 & $ k^{5-\eta}$ & \phantom{-}1 & $\tilde{f}_0+\tilde{f}_2 k^2$                   \\
5 & $ k^{6-\eta}$ & \phantom{-}2 & $\tilde{f}_0+\tilde{f}_2 k^2 + \tilde{f}_4 k^4$ \\ 
6 & $ k^{7-\eta}$ & \phantom{-}2 & $\tilde{f}_0+\tilde{f}_2 k^2 + \tilde{f}_4 k^4$ \\ 
$\vdots$ & $\vdots$ & $\vdots$ & $\vdots$ \\
\bottomrule[0.95pt]
%\hline
\end{tabular}
\end{center}
\caption{Counterterms expansion of the moments of first orders.}
\label{tab:counterterms}
\end{table}
The regularization procedure ensures the convergence of the integrand in Eq.~\ref{sect2eq08} over the integration domain. For values of $m$ close to even integers, the logarithmic divergence of the integral is balanced by the vanishing $\mathcal{N}_{\lambda}$ to give a finite quantity. More precisely, considering $(r^{m-\eta},f)$ for even $m$=$2p$, the normalization coefficient $\mathcal{N}_{2p-\eta}$ in the vicinity of $\eta=0^+$ can be written as
\begin{equation}\label{Neta}
\mathcal{N}_{2p - \eta} \simeq {(-1)}^{ p} \, (2p + 1)! \, \eta \, .
\end{equation}
Introducing an intermediate momentum $Q$, the integral of Eq.~\ref{sect2eq08} can be separated into a contribution dominated by the zero-momentum behaviour of the integrand and another depending on its infinite momentum behaviour. In the vicinity of zero-momentum, the integrand behaves as $\tilde{f}_{2p} / k^{1-\eta}$ leading, after $k$-integration, to the contribution $\tilde{f}_{2p} Q^{\eta}/\eta$. At large momentum, the $k$-dependence of the integrand ensures a finite $I_Q$ value for the infinite momentum integral. Then, even moments can be recast as 
\begin{eqnarray}
( r^{2p},f ) & = & \lim_{\eta \to 0^+} {(-1)}^{ p} \, (2p + 1)! \, \eta  \, \left( \frac{ \tilde{f}_{2p}}{\eta} Q^{\eta} + I_Q \right) \nonumber \\
& = & {(-1)}^{ p} \, (2p + 1)! \, \tilde f_{2p} \, . \label{MomRedEve}
\end{eqnarray} 
For instance, we have $(r^0,f) = \tilde{f}_0$, $(r^2,f)=-6\tilde{f}_2$, $(r^4,f)=120 \tilde{f}_4$... as expected from the MacLaurin development of the Fourier transform $\tilde{f}(\bold{k})$.

The regularization of the Fourier transform $g_{\lambda}(\bold{k})$ of the tempered distribution $r^{\lambda}$ is not unique. For instance, $g_{\lambda}(\bold{k})$ can also be given as a weak limit of the convergent integral
\begin{eqnarray}\label{sect2eq03p}
g_{\lambda}(\bold{k}) = \lim_{\epsilon \to 0^+} \int_{\nbR^3} \mathrm{d}^3 \bold{r} \, r ^{\lambda} e^{-\epsilon r} \, e^{i \, \bold{k}\cdot \bold{r} } = \lim_{\epsilon \to 0^+} {\mathcal I}_{\lambda}(k,\epsilon)
\end{eqnarray}
where the term $e^{-\epsilon r}$ ensures the convergence of the integral ${\mathcal I}_{\lambda}(k,\epsilon)$. This is a standard technique used, for example, to regularize the Fourier transform of the Coulomb potential~\cite{Fet80,Alt19}. The integration of Eq.~\ref{sect2eq03p} is analytical and yields for any $\lambda > -3$ and $\lambda \neq -2$
\begin{equation}
{\mathcal I}_{\lambda}(k,\epsilon) = \frac{4 \pi \, \Gamma(\lambda+2) \, \sin\left[ (\lambda+2) {\rm Arctan} \left( k/\epsilon \right) \right]}{k (k^2 + \epsilon^2)^{\frac{\lambda}{2}+1} } 
\end{equation}
which accepts the limit $(4 \pi/k) {\rm Arctan} \left( k/\epsilon \right)$ at $\lambda$=$-2$. The moments defined in Eq.~\ref{sect2eq02} can then be written as
\begin{eqnarray}\label{sect2eq05p}
( r^{\lambda}, f ) & = & \frac{2}{\pi}  \, \Gamma(\lambda+2) \, \times \\
& & \lim_{\epsilon \to 0^+} \int_{0}^{\infty} \mathrm{d}k \, \tilde{f}(k) \, \frac{k \sin\left[ (\lambda+2) {\rm Arctan} \left( k/\epsilon \right) \right]}{(k^2 + \epsilon^2)^{\lambda/2+1}} \nonumber
\end{eqnarray}
for any $\lambda > -3$ and $\lambda \neq -2$ value. For integer values of $\lambda$, the sine function in Eq.~\ref{sect2eq05p} can be developed in terms of a $k/\epsilon$ polynomial, such that Eq.~\ref{sect2eq05p} can be recast for $\lambda$=$m$ as
\begin{eqnarray}
( r^m, f ) & = & \frac{2}{\pi} \, (m+1)! \, \times \label{ImAna} \\
& & \lim_{\epsilon \to 0^+} \epsilon^{m+2} \int_0^{\infty} \mathrm{d}k \, \tilde{f}(k) \, \frac{k}{ (k^2 + \epsilon^2)^{m+2} } \, \Phi_{m}(k/\epsilon) \nonumber
\end{eqnarray}
with
\begin{equation} \label{eqfim}
\Phi_{m}(k/\epsilon) =\sum_{j=0}^{m+2} \sin \left( \frac{j \pi}{2} \right) \frac{(m+2)!}{j!(m+2-j)!} \, \left( \frac{k}{\epsilon} \right)^j \, .
\end{equation}

The formulations of Eq.~\ref{sect2eq08} and Eq.~\ref{sect2eq05p} allow us to determine the moments of a given operator directly in the momentum space, for both integer and non-integer values of $\lambda$. For a given $\tilde{f}(k)$ functional form, the moments are numerically computed from these expressions and can also be obtained analytically for specific cases.

%
% ------------------------------------------------------------------------------------------------------------------------
%
\section{Applicability and benefit of the integral method} \label{sec:meth}

The momentum integral determination of the moments outlined in the previous section is a general approach that can be applied to any relevant physics quantity. Without any restriction on the applicability of the method, the specific case of the electromagnetic form factors of the proton is considered hereafter. A typical function example is the radial density
\begin{equation} \label{dip-den}
f_D( \bold{r} ) = \frac{\Lambda^3}{8\pi} \, e^{-\Lambda r}
\end{equation}
leading to the well-known dipole parameterization
\begin{equation}\label{ffdip}
\tilde{f}_D(\bold{k}) = \int_{\nbR^3} \mathrm{d}^3\bold{r} \, e^{- i \bold{k}\cdot \bold{r}} f_D(\bold{r}) = \frac{\Lambda^4}{(k^2+\Lambda^2)^2} 
\end{equation}
where $\Lambda$ represents the dipole mass parameter. The moments can be determined directly in the configuration space, as 
\begin{equation} \label{dip-mom}
(r^{\lambda},f_D) = \int_{\nbR^3} \mathrm{d}^3\bold{r} \, r^{\lambda} \, f_D(\bold{r}) = \frac{\Gamma(\lambda+3)}{2} \, \frac{1}{\Lambda^{\lambda}} \, .
\end{equation}
Considering integer $\lambda$=$m$ values, Eq.~\ref{ImAna} can be written as
\begin{equation}\label{jmint}
(r^{m},f_D) = \frac{2 \, \Gamma(m+2)}{\pi} \, \frac{1}{\Lambda^m} \, \lim_{\tilde{\epsilon} \to 0^+} J_m(\tilde{\epsilon})
\end{equation}
with $\tilde{\epsilon}$=$\epsilon/\Lambda$, and from Eq.~\ref{ImAna} with the integral variable change $z$=$k/\epsilon$
\begin{eqnarray} \label{intres}
J_m(\tilde{\epsilon}) & = & \frac{1}{\tilde{\epsilon}^m} \sum_{j=0}^{m+2} \sin \left( \frac{j \pi}{2} \right) \frac{(m+2)!}{j!(m+2-j)!} \, \times \\
& & \int_0^{\infty} \mathrm{d}z \, \frac{z^{j+1}}{(1+\tilde{\epsilon}^2z^2)^2(1+z^2)^{m+2}} = \frac{\pi}{4} \frac{m+2}{(1+\tilde{\epsilon})^3} \, . \nonumber
\end{eqnarray}
Evaluating the limit in Eq.~\ref{jmint}, the momentum integral expression of the moments becomes
\begin{equation}
(r^{m},f_D) = \frac{2 \, \Gamma(m+2)}{\pi \Lambda^m} \frac{\pi (m+2)}{4} = \frac{\Gamma(m+3)}{2} \frac{1}{\Lambda^{m}} \, \label{rmfD}
\end{equation}
{\it i.e.} identical to the result of Eq.~\ref{dip-mom} obtained from the configuration space integral. The same result is obtained for any real (integer and non-integer) $\lambda$ value from the numerical evaluation of the integrals in Eq.~\ref{sect2eq08} and Eq.~\ref{sect2eq05p}. The method has been tested for different mathematical realizations of the radial function $f(\bold r)$ and several $\lambda$: the exponential form of Eq.~\ref{dip-den}, and a Yukawa-like form (see~\ref{App1}) corresponding to the parameterization of the proton electromagnetic form factors in terms of a $k^2$-polynomial ratio, the Kelly's parameterization~\cite{Kel04}. In each case, the numerical evaluation of Eq.~\ref{sect2eq08} and Eq.~\ref{sect2eq05p} provides with a very high accuracy the same results as the configuration space integrals.

\begin{figure} [t!]
\begin{center}
\includegraphics[width=0.995\columnwidth]{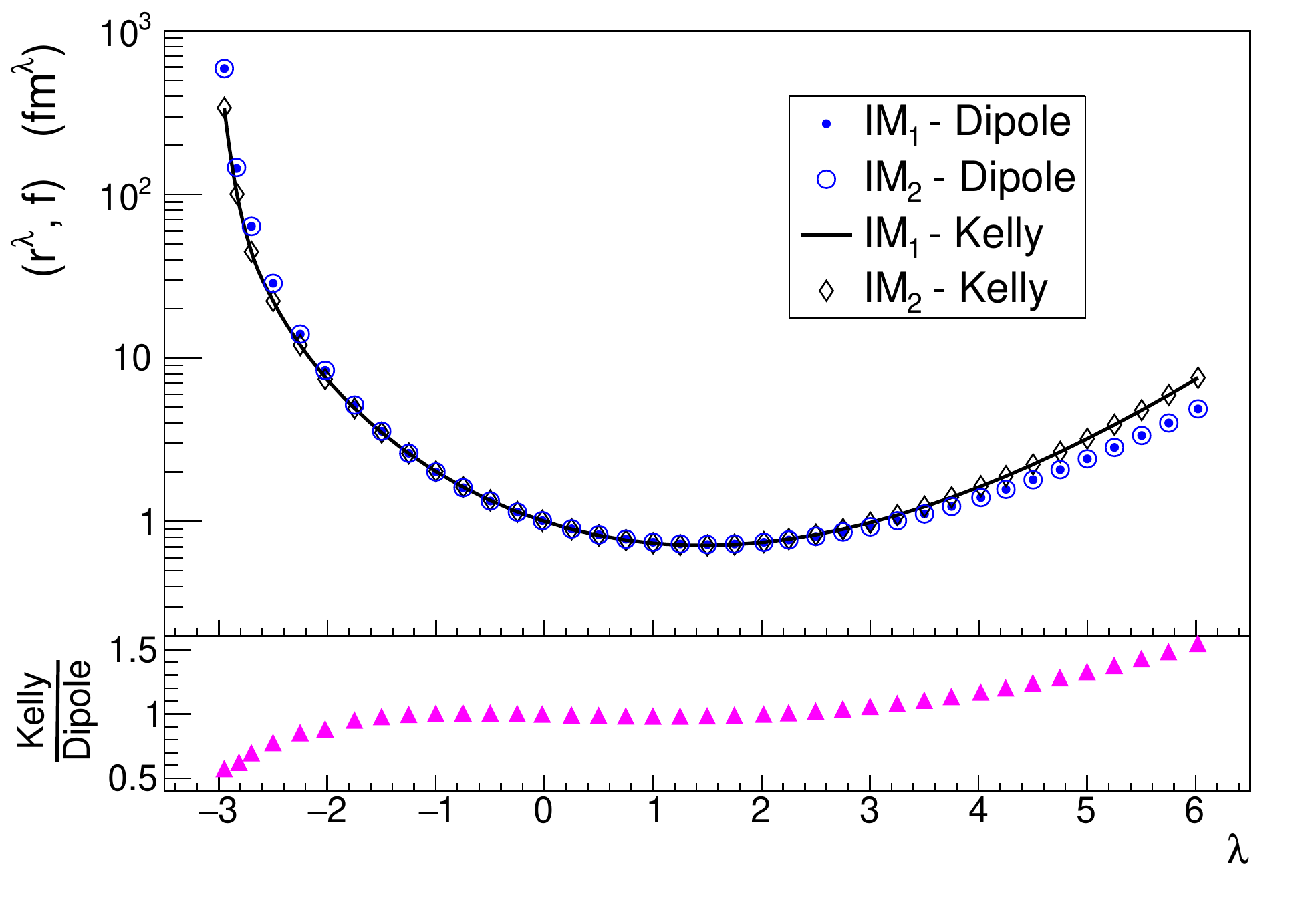}
\caption{$\lambda$-order moments of the proton electric form factor, determined from the integral method for the dipole  ($\Lambda^2$=16.1~fm$^{-2}$) and the Kelly's polynomial ratio~\cite{Kel04} parameterizations (top panel), and ratio between the two parameterizations (bottom panel).} 
\label{fig:inf_mom}
\end{center}
\end{figure}
Figure~\ref{fig:inf_mom} shows the variation of the moments over a selected $\lambda$-range for two parameterizations of the electric form factor of the proton and both prescriptions of the integral method: the principal value regularization of Eq.~\ref{sect2eq08} denoted IM$_1$, and the exponential regularization of Eq.~\ref{sect2eq05p} denoted IM$_2$. Particularly, the two different numerical evaluations are shown to deliver, as expected, exactly the same results (top panel of Fig.~\ref{fig:inf_mom}). Because of a similar functional form, the polynomial ratio moments do not strongly differ from the dipole moments. Nevertheless, sizeable differences can be observed for negative $\lambda$'s and high moment orders (bottom panel of Fig.~\ref{fig:inf_mom}). Negative orders are relevant for the study of the high-momentum dependence of the form factor ({\it i.e.} the central part of the corresponding density), and are of interest to probe its asymptotic behaviour, whereas the high positive order moments probe the low-momentum behaviour of the form factor (namely the density close to the nucleon's surface). 

%
% ------------------------------------------------------------------------------------------------------------------------
%
\section{Application to experimental data} \label{sec:app}

The integral method described previously relies on integrals of Fourier transforms {\it i.e.} form factors for the present discussion. Unlike the derivative method, the integral method is less sensitive to a very small variation of the form factor at low momentum, and a more stable behaviour with respect to the functional form can be expected. However, the evaluation of moments via this method requires an experimentally defined asymptotic limit which may be hardly obtained considering the  momentum coverage of actual experimental data. The momentum dependence of the integrands of Eq.~\ref{sect2eq08} and Eq.~\ref{sect2eq05p} provides the solution to this issue. The denominator of the integrands scales at large momentum like $k^{\lambda+1}$, meaning that the integrals are most likely to saturate at a momentum value well below infinity. 

Truncated moments, defined from Eq.~\ref{sect2eq08} and Eq.~\ref{sect2eq05p} by replacing the infinite integral boundary by a cut-off $Q$, allow us to understand the saturation behaviour of the moments. Considering for sake of simplicity the case of integer $\lambda$=$m$ values, they can be written from Eq.~\ref{ImAna}
\be
(r^m,f)_Q = \frac{2}{\pi} \, (m+1)! \, { \lim_{\epsilon\to 0^+} {\mathcal R}_m(Q,\epsilon) } \label{eq:rmQ}
\ee
with
\be
{\mathcal R}_m(Q,\epsilon) = \epsilon^{m+2} \, \int_0^Q \mathrm{d}k \, \tilde{f}(k) \, \frac{k \, \Phi_m(k/\epsilon)}{(k^2+\epsilon^2)^{m+2}} \, . \label{Rinte}
\ee
The integral is performed before taking the $\epsilon$-limit, and obviously 
\be
\lim_{Q \to \infty} (r^m,f)_{_Q} \, = \, (r^m,f) \, . \label{eq:mominf}
\ee
For the typical example of the dipole parameterization of Eq.~\ref{ffdip}, the integral for even and odd moments can be expressed as  
\ba
{\mathcal R}_{2p}(Q,\epsilon) & = & \epsilon \, u_{2p}(Q,\epsilon) + \epsilon \, v_{2p}(\epsilon) \, {\rm Arctan} \left( \frac{Q}{\Lambda} \right) \nonumber \\
& + & w_{2p}(\epsilon) \, {\rm Arctan} \left( \frac{Q}{\epsilon} \right) \label{Jev} \\
{\mathcal R}_{2p+1}(Q,\epsilon) & = & u_{2p+1}(Q,\epsilon) + v_{2p+1}(\epsilon) \, {\rm Arctan} \left( \frac{Q}{\Lambda} \right) \nonumber \\
 & + & \epsilon \, w_{2p+1}(\epsilon) \, {\rm Arctan} \left( \frac{Q}{\epsilon} \right) \, . \label{Jod}
\ea
The functions $u_i$'s, $v_i$'s, and $w_i$'s have finite limits when $\epsilon \to 0^+$, as well as when $Q \to \infty$ for the $u_i$'s. Moreover, the $v_i$'s and $w_i$'s are independent of $Q$. The structure of Eq.~\ref{Jev} and Eq.~\ref{Jod}  exhibits three contributions with different $Q$-dependences: the first term (with $u_i$'s) corresponds to a ratio of $Q$-polynomials and vanishes as $1/Q$ at infinite cut-off; the second term (with $v_i$'s) varies as ${\rm Arctan}(Q/\Lambda)$ and is related to the $k_0$=$\pm i\Lambda$ complex pole of the $\tilde{f}_D({\bold k})$ function; the last term (with $w_i$'s) saturates as ${\rm Arctan}(Q/\epsilon)$  and is associated to the $k_0$=$\pm i\epsilon$ complex pole of the function that samples $\tilde{f}_D({\bold k})$. The $Q$-convergence of the two last terms is determined by the same asymptotic behaviour
\be
\lim_{x \to +\infty}{\rm Arctan} (x) = \frac{\pi}{2} - \frac{1}{x} + \frac{1}{3x^3} + {\mathcal O}\left( \frac{1}{x^5}\right) \, .
\ee
The $\epsilon$ factor in front of these contributions distinguishes the saturation behaviour of even and odd moments. Particularly, in the limit $\epsilon \to 0^+$, the even truncated moments write
\be
( r^{2p},f_D )_Q = (2p+1)! \, w_{2p}(0^+) = \frac{(2p+2)!}{2} \, \frac{1}{\Lambda^{2p}} \label{eq:eve0}
\ee
and are independent of $Q$, while the odd truncated moments 
\ba
& & (r^{2p+1},f_D)_Q = \frac{2}{\pi} \, {(2p+2)!} \, \times \\
& & \left[ u_{2p+1}(Q,0^+) + v_{2p+1}(0^+) \, {\rm Arctan} \left(\frac{Q}{\Lambda}\right) \right] \nonumber \label{eq:odd0}
\ea
are still depending on the cut-off. Indeed, Eq.~\ref{Jev} can be seen as a different realization of Eq.~\ref{MomRedEve}, similarly leading to the $Q$-independence of even moments. The $u_i$'s coefficients behave like $1/Q$ functions at large cut-off, and consequently vanish for infinite $Q$. For example, the first odd coefficients write
\ba
u_1(Q,0^+) & = & \frac{2 \Lambda^2 + 3 Q^2}{2 Q \left(\Lambda^2+Q^2\right)} \xrightarrow[Q\to \infty]{} 0 \\
  v_1(0^+) & = & \frac{3}{2 \Lambda} \\
u_3(Q,0^+) & = & \frac{-2 \Lambda^4+10 \Lambda^2 Q^2+15 Q^4}{6 \Lambda^2 Q^3 \left(\Lambda^2+Q^2\right)} \xrightarrow[Q\to \infty]{} 0 \\
  v_3(0^+) & = & \frac{5}{2\Lambda^3} \, .
\ea
Only the $v_i$'s remain in the infinite $Q$-limit, leading to the expression of Eq.~\ref{rmfD}. Similar features are derived in~\ref{App2} for the Kelly's parameterization.

\begin{figure} [t!]
\begin{center}
\includegraphics[width=0.995\columnwidth]{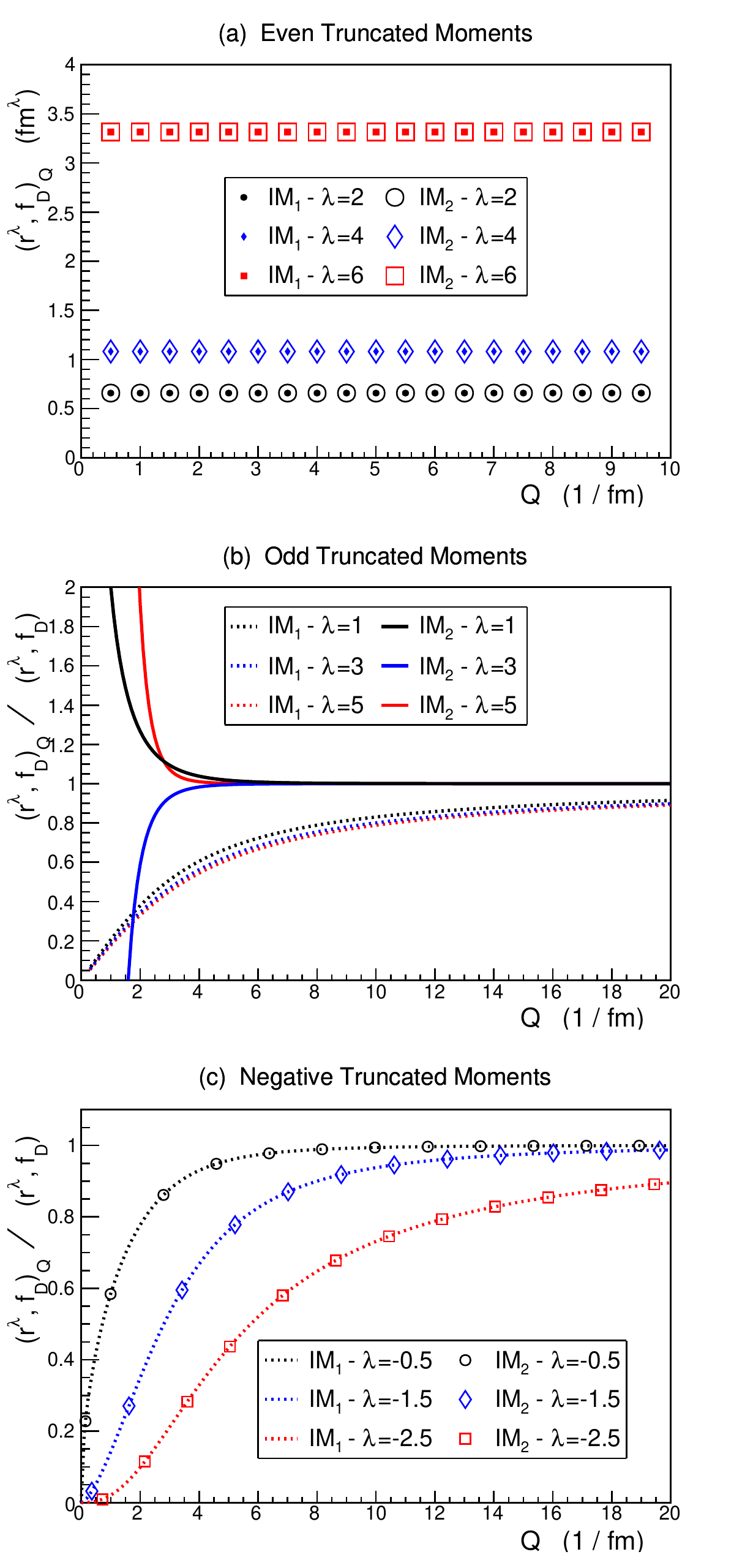}
\caption{Convergence of truncated moments of the proton electric form factor for selected orders within the dipole parameterization: (a) positive even, (b) positive odd, and (c) negative non-integer. IM$_1$ and IM$_2$ denote the principal value and the exponential regularizations, respectively.}
\label{fig:Qsat} 
\end{center}
\end{figure}
The $Q$-convergence of truncated moments is shown in Fig.~\ref{fig:Qsat} for selected moment orders, as determined for the two prescriptions of the integral method (IM$_1$ and IM$_2$) where the $Q$ cut-off replaces the infinite boundary of the integrals. The $Q$-independence feature of even truncated moments is reproduced by each prescription (Fig.~\ref{fig:Qsat}(a)). This is a general feature independent of the specific form factor, as expressed by Eq.~\ref{MomRedEve}. In other words, the integral method for even moments recovers formally the same quantities as the derivative method. In the ideal world of perfect experiments, adjusting experimental data with the same function over a small or large $k^2$-domain affects only the precision on the parameters of the function. In the context of the limited quality of real data, the integral method provides the mathematical support required to consider the full $k^2$-unlimited domain of existing data, leading therefore to a more accurate determination of the moments. The practical constraint is to obtain an appropriate description of the data over a large $k^2$-domain. \newline
Fig.~\ref{fig:Qsat}(b) shows the $Q$-convergence of selected odd moment, comparing the integral method prescriptions. The different regularizations of the $g_{\lambda}({\bold k})$ integral lead to different saturation behaviours. While the principal value regularization (IM$_1$) asks for large $Q$-values, the exponential regularization (IM$_2$) rapidly saturates about $6$~fm$^{-1}$, {\it i.e.} in a momentum region well covered by proton electromagnetic form factors data~\cite{Pun15}. \newline
Fig.~\ref{fig:Qsat}(c) shows the $Q$-convergence of selected moments with negative non-integer orders. For such orders, there are no counterterms for the principal value regularization (Tab.~\ref{tab:counterterms}), and the effect of the exponential regularization term in Eq.~\ref{sect2eq03p} is strongly suppressed since the integrand converges at infinity (for $-3<\lambda<-1$). Indeed, there is no need of regularization for negative orders and all prescriptions of the integral method should be identical. This is verified on Fig.~\ref{fig:Qsat} where the numerical evaluation of each prescription is shown to provide the same result: IM$_1$=IM$_2$ for $-3<\lambda<0$. \newline 
It is the essential  benefit of the integral method to allow us to determine odd and real positive and negative spatial moments directly from experimental data in the momentum space.
\begin{figure}[t!]
\begin{center}
\includegraphics[width=0.995\columnwidth]{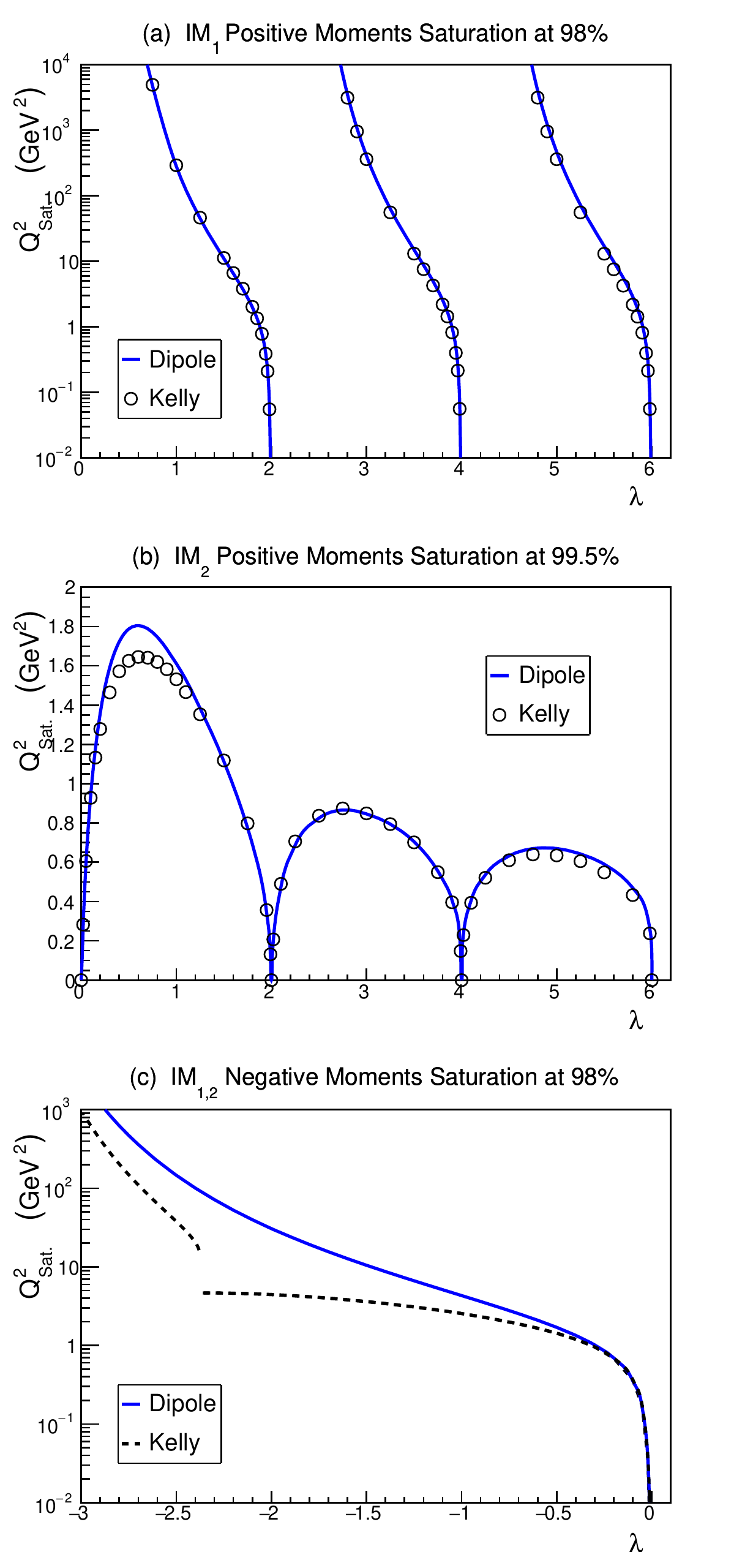}
\caption{Saturation momentum of the principal value (IM$_1$) and exponential (IM$_2$) regularizations of the integral method, for the dipole (solid line) and Kelly~\cite{Kel04} (circle and dashed line) parameterizations of the electric form factor of the proton: (a) 98\% saturation of positive moments within the IM$_1$ prescription, (b) 99.5\% saturation of positive moments within the IM$_2$ prescription, and (c) 98\% saturation of negative  moments. The latter is independent of the integral method prescription.}
\label{fig:Qswe} 
\end{center}
\end{figure}

We define the saturation momentum $Q_{Sat.}$ for each moment order as the squared momentum transfer at which the truncated moment is some  $\alpha$-fraction of the true moment value obtained in the limit $Q \to \infty$ (Eq.~\ref{eq:mominf}), that is
\be
R^{\lambda}_{Q_{Sat.}} = \frac{(r^{\lambda},f)_{Q_{Sat.}}}{(r^{\lambda},f)} = \alpha \, .
\ee
The variation of the saturation momentum as a function of the moment order is shown on Fig.~\ref{fig:Qswe} for both prescriptions of the integral method and two parameterizations of the electric form factor of the proton. The 98\% saturation ($\alpha$=~0.98) of IM$_1$ (Fig.~\ref{fig:Qswe}(a)) is compared to the 99.5\% saturation of IM$_2$ (Fig.~\ref{fig:Qswe}(b)), with respect to positive moments. The principal value regularization appears less performant than the exponential regularization. The differences between the integrands of each prescription is responsible for this behaviour. At a maximum squared momentum transfer of 2~GeV$^2$, the IM$_2$ prescription permits the determination of any positive moments, while the IM$_1$ prescription is of very limited success, even when considering a less demanding saturation and the full extension of the $k^2$-domain of existing data up to $\sim$10~GeV$^2$. Noticeably, the saturation momentum appears weakly dependent on the form factor model (Fig.~\ref{fig:Qswe}(a) and (b)). \newline 
Negative moments are more difficult to obtain very accurately but can still be determined with a few percents precision  (Fig.~\ref{fig:Qswe}(c)). The sensitivity to the form factor parameterization is particularly remarkable. As noted previously in Sec.~\ref{sec:meth}, negative moments are sensitive to the high-momentum behaviour of the form factor which 
is only partly covered by actual data. Here, the difference of interest between the parameterizations is the sign change of $G_E(k^2)$ predicted at $k^2_0$=14.7~GeV$^2$ in Kelly's. This results in a maximum ratio value at $k^2$ such that $R^{\lambda}_{k_0}>1$, and provides a saturation momentum $Q_{Sat.} < k_0$ ($Q_{Sat.} > k_0$) when $R^{\lambda}_{k_0}<2-\alpha$ ($R^{\lambda}_{k_0}>2-\alpha$). These two regimes are responsible for the discontinuity occuring about $\lambda$=-2.4 in Fig.~\ref{fig:Qswe}(c). Note that the moment order corresponding to the discontinuity is not a constant but depends on the $\alpha$ saturation level. Negative moments clearly magnify the impact of the change of the sign of the form factor, and may be used to discriminate different form factor models.

\begin{figure}[t!]
\begin{center}
\includegraphics[width=0.995\columnwidth]{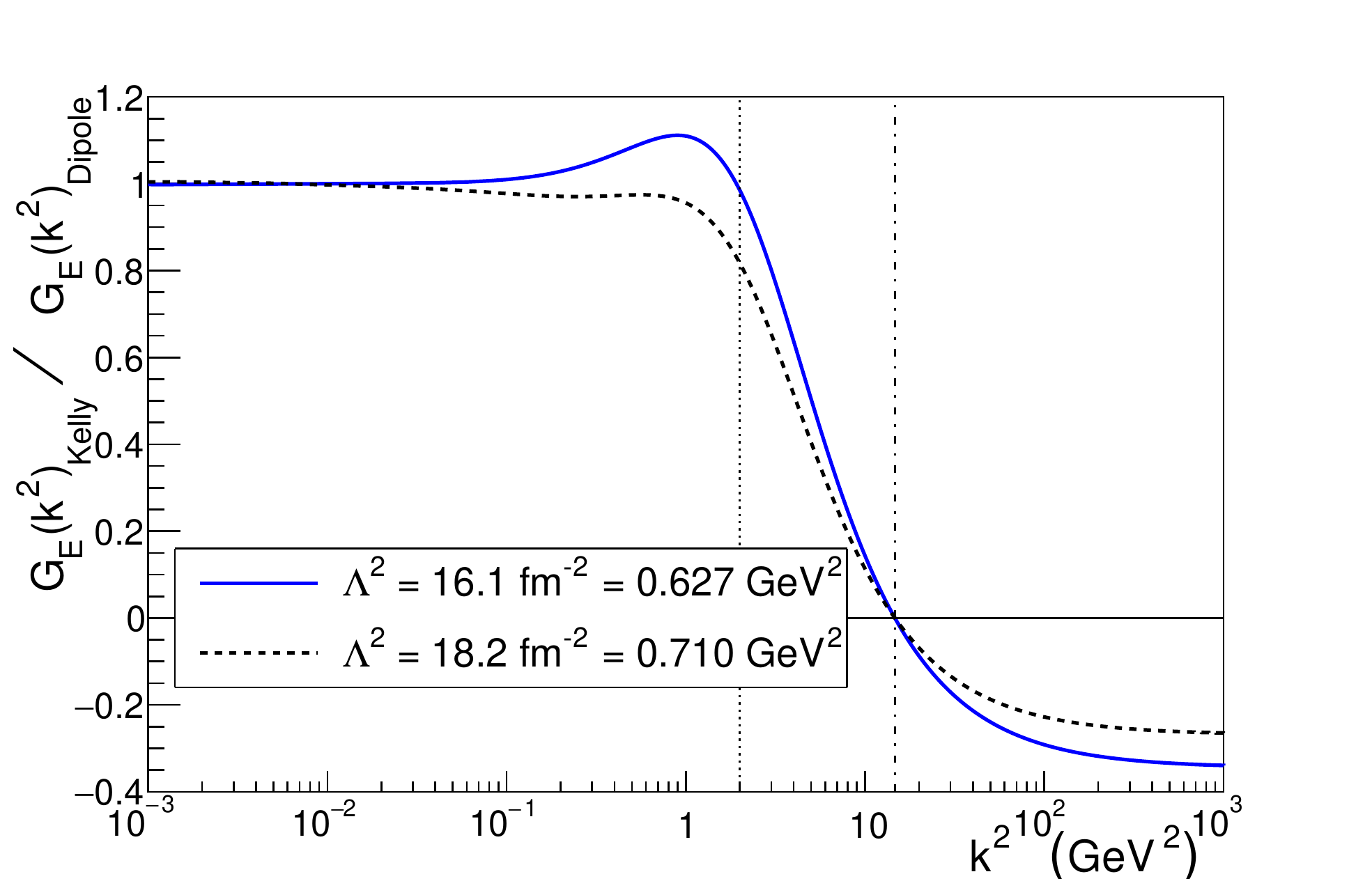}
\caption{Kelly parameterization of the electric form factor of the proton normalized by the dipole parameterization for different dipole masses: the mass used in the present work (solid line), and the historical parameterization mass (dashed line). The saturation momentum at 2~GeV$^2$ (vertical dotted line) and the zero-crossing momentum (vertical dash-dotted line) are also shown.}
\label{fig:keldip} 
\end{center}
\end{figure}

A closer look at the form factor parameterizations explains further Fig.~\ref{fig:Qswe} behaviours. The $k^2$-dependences of the electric form factor of the proton within the Kelly and the dipole parameterizations are compared in Fig.~\ref{fig:keldip} for two different dipole masses. Up to the momentum saturation of 2~GeV$^2$, the differences between the parameterizations are small ($\sim$10\% at most), which leads to the very similar saturation momentum behaviour observed for moments of positive orders (Fig.~\ref{fig:Qswe}). More precisely, the Kelly's moments differ from the dipole ones (Fig.~\ref{fig:inf_mom}) but both kinds converge similarly towards the asymptotic limit. Differences only show up for the lowest order moments (Fig.~\ref{fig:Qswe}(b)) which succeed to catch changes in the $k^2$-dependences above $\sim$1~GeV$^2$. In the region between the saturation momentum and the zero-crossing momentum, the parameterizations strongly differ in magnitudes and $k^2$-dependences  (Fig.~\ref{fig:keldip}). This leads to the very different saturation momentum trends observed in the moment region -$2.4<\lambda \le 0$ in Fig.~\ref{fig:Qswe}(c). When the moment order is large enough (-$3<\lambda<\,$-2.4) to sample the high-$k^2$ region of the form factor where the parameterizations have identical $k^2$-dependences (Fig.~\ref{fig:keldip}), the behaviours of the saturation momentum become similar (Fig.~\ref{fig:Qswe}(c)).

These features remain model-dependent in the sense that the high-momentum behaviour of the form factors is deduced from predicted scaling  laws~\cite{Bro73} which, because of the limited experimental knowledge, are not confirmed by existing data. However, the momentum range spanned by actual data, especially for the proton, is large enough to sufficiently constrain any physical or phenomenological parameterization. Therefore a momentum saturation quasi-independent of the functional realization of the proton form factor can be determined for positive moments. Major differences attached to the high-momentum region are specifically showing up for negative moments. 

%
% ------------------------------------------------------------------------------------------------------------------------
%
\section{Conclusions} \label{sec:concl}

The present work proposes a new method to determine the spatial moments of densities expressed in the momentum space, {\it i.e.} form factors. The method provides a direct access to real moments, both positive and negative, for any form factor functional. Particularly, it represents the only opportunity to access spatial moments when the Fourier transform of a parameterization cannot be performed. In addition, unlike the derivative method which is restricted to even moments, the so-called integral method gives access to any moment order, especially odd moments and more generally any real moment with $\lambda > -3$. Furthermore, it provides the formal support to take into account the full range of existing data for the determination of even moments, allowing us to improve their accuracy as compared to the derivative method.

The integral method involves the regularization of integrals treated as distributions. Two regularization schemes were studied: the first one based on the principal value regularization, similar to the technique used to determine Zemach moments~\cite{Zem56}; the second one involving an exponential regularization, similar to the technique used to regularize the Fourier transform of the Coulomb potential~\cite{Fet80}. These techniques have been tested with respect to the dipole and Kelly parameterizations of the electromagnetic form factor of the proton. The exponential regularization provides the most performant approach allowing us to determine accurately positive moments considering a squared saturation four momentum transfer of 2~GeV$^2$. Negative moments require larger saturation momenta but remain quite accesssible with reduced accuray (a few percents) in the proton case. \newline
The integral method is not specific of the proton, and can also be applied to the neutron and nuclei electromagnetic form factors. These applications will be presented elsewhere.
 
%
% ------------------------------------------------------------------------------------------------------------------------
%
\section*{Acknowledgements}

This work was supported by the LabEx Physique des 2 Infinis et des Origines (ANR-10-LABX-0038) in the framework {$\ll$~Investissements d'Avenir~$\gg$} (ANR-11-IDEX-01), the French Ile-de-France region within the SESAME framework, the INFN under the Project Iniziativa Specifica MANYBODY, and the University of Turin under the Project BARM-RILO-19. This project has received funding from the European Unions's Horizon 2020 research and innovation programme under grant agreement No 824093.

%
% ------------------------------------------------------------------------------------------------------------------------
%
\appendix

\section{Partial waves expansion of radial moments} \label{App0}

This appendix demonstrates that only the spherical components of the form factor $f(\bold{r})$ contribute to the radial moments defined in Eq.~\ref{sect2eq01}.

Consider any real number $\lambda$ and any function $ f(\bold{r}) $  of the three-dimensional variable $\bold{r}$, and further assume that the integral defined as
\be\label{eq01}
I_{\lambda} =  \int_{\mathrm{I\! R^3}  } f(\bold{r}) \, r^{\lambda} \ d^3 \bold{r}
\ee
is finite. Any function $f(\bold{r})$ can be expanded in partial waves as follows 
\be\label{eq02}
f(\bold{r}) = \sum_{\ell=0}^\infty  \sum_{m=-\ell}^\ell  \ \beta_{\ell m} (r ) \ Y_{\ell m}^{*} ( \hat{\bold{r}} ) 
\ee
with 
\be\label{eq03}
 \beta_{\ell m} (r) = \int f(\bold{r}) \ Y_{\ell m}( \hat{\bold{r}} ) \ d\hat{\bold{r}} \, , 
\ee
such that 
\ba
I_{\lambda} = \sum_{\ell=0}^\infty  \sum_{m=-\ell}^\ell  \int \beta_{\ell m} (r ) \ r^{2  + \lambda} \ Y_{\ell m}^{*} ( \hat{\bold{r}} ) \ d\hat{\bold{r}} dr \, .
\ea 
Using
\be
\int Y_{\ell m}^{*} ( \hat{\bold{r}} )  \  d\hat{\bold{r}} = \sqrt{4\pi} \,  \delta_{\ell 0} \, \delta_{m 0} 
\ee
we obtain
\be
 I_{\lambda} =  \sum_{\ell=0}^\infty \sum_{m=-\ell}^\ell \  \left[I_{\lambda} \right]_{\ell m}  =  \left[I_{\lambda} \right]_{00} 
\ee
where
\be
\left[I_{\lambda} \right]_{00} = \int_0^{\infty}  \beta_{00} (r ) \ r^{2  + \lambda} \, dr \,.
\ee
Therefore, $I_{\lambda}$ vanishes for any $\ell \neq 0$, {\it i.e.} only the partial wave $\ell$=$0$ contributes to the integral. Consequently, any pure radial function or any function whose partial wave expansion have a spherical ($\ell$=$0$) term lead to a non-vanishing $I_{\lambda}$. Moreover, the Fourier transform of this spherical part will be induced only by the $j_0(kr)$ spherical Bessel function.

%
% ------------------------------------------------------------------------------------------------------------------------
%
\section{Moments of a polynomial ratio form factor} \label{App1}

This appendix discusses the determination in the configuration space of the moments of a function having Fourier transform in momentum space expressed as a polynomial ratio. These results serve the comparison with the moments obtained in Sec.~\ref{sec:meth} from the momentum integral method.

Considering the polynomial ratio function $\tilde{f}_K(\bold k)$ expressed in momentum space as 
\begin{equation}\label{ffkel}
\tilde{f}_K(\bold k) \equiv \tilde{f}_K(k) = \frac{1+a_1k^2}{1+b_1k^2+b_2k^4+b_3k^6} \, ,
\end{equation}
its inverse Fourier transform writes
\begin{equation}
f_K(\bold r) \equiv f_K(r) = \frac{1}{2\pi^2} \, \frac{1}{r} \, \int_0^{\infty} \mathrm{d}k \, k \tilde{f}_K(k) \, \sin(kr) \, .
\end{equation}
$\tilde{f}_K(\bold k)$ is assumed to represent a regular physics quantity, for instance the electromagnetic form factors of the nucleon~\cite{Kel04}, such that the denominator never vanishes for real $k$ and the function accepts only complex poles. The product $k \tilde{f}_K(k)$ can then be expanded in partial fractions as
\begin{equation}
k \tilde{f}_K (k) = \sum_{i=1}^3 \left[ \frac{A_i}{k-k_i} + \frac{\overline{A_i}}{k-\overline{k_i}} \right]
\end{equation}
where the $k_i$'s (with $\Im{\rm m}[k_i]>0$) are the poles of $\tilde{f}(k)$, and
\be
A_i = - \frac{i}{2b_3} \, \frac{(1+a_1k_i^2) k_i}{\Im{\rm m}[k_i]} \bigg{/} \prod_{j(\ne i)=1}^3 (k_i-k_j)(k_i-\overline{k_j}) \, .
\ee
are the residues of the function $k \tilde{f}_K (k)$ at $k$=$k_i$.The numerical values of the $A_i$'s and $k_i$'s corresponding to the parameterization of Ref.~\cite{Kel04} for the electric and magnetic proton form factors are listed in Tab.~\ref{Gmpoles}. After integration, the radial function writes
\begin{eqnarray}
 f_K(\bold r) & = & \frac{1}{2\pi} \, \frac{1}{r} \, \sum_{i=1}^3 e^{-\Im{\rm m}[k_i] r}  \, \times \\
& & \bigg[ \Re{\rm e}[A_i] \cos{\big(\Re{\rm e}[k_i] r \big)} - \Im{\rm m}[A_i] \sin{\big( \Re{\rm e}[k_i] r \big)} \bigg] \, . \nonumber
\end{eqnarray}
The absence of odd powers of $k$ in the denominator of $\tilde{f}_K({\bold k})$ leads to the relationships
\be \sum_{i=1}^3 \Re{\rm e}[A_i] = \sum_{i=1}^3 \Re{\rm e}[k_i] = 0 \label{relaKel} \ee
which ensure a finite value of $f_K(\bold r)$ at $r$=$0$. The moments, determined from the configuration space integral of Eq.~\ref{sect2eq01}, can be expressed as
\begin{eqnarray}
( r^{\lambda}, f_K ) & = & 2 \, \Gamma(\lambda+2) \, \times \\
& & \sum_{i=1}^3 \frac{ \Re{\rm e}[A_i] \cos(\theta_{k_i}) - \Im{\rm m}[A_i] \sin(\theta_{k_i}) }{{\vert k_i \vert}^{\lambda+2}} \nonumber
\end{eqnarray}
with $\lambda > -2$ and 
\be
\theta_{k_i} = (\lambda+2) \, {\rm Arctan} \left( \frac{\Re{\rm e}[k_i]}{\Im{\rm m}[k_i]} \right) \, .
\ee

\begin{table}[t!]
\ra{1.1}
\begin{center}
\begin{tabular}{@{}ccccccccc@{}}
\toprule[0.95pt]
    & \multicolumn{4}{c}{$\pmb{G_{E_p}}$} & \multicolumn{4}{c}{$\pmb{G_{M_p}}\pmb{/\mu_p}$} \\ 
    %\cline{2-5}\cline{6-9}
    \cmidrule[0.5pt](lr){2-5}\cmidrule[0.5pt](lr){6-9}
$\pmb{i}$ & \multicolumn{2}{c}{$\pmb{k_i}$ \bfseries{(fm$^{\pmb{-1}}$)}} & \multicolumn{2}{c}{$\pmb{A_i}$ \bfseries{(fm$^{\pmb{-2}}$)}} & \multicolumn{2}{c}{$\pmb{k_i}$ \bfseries{(fm$^{\pmb{-1}}$)}} & \multicolumn{2}{c}{$\pmb{A_i}$ \bfseries{(fm$^{\pmb{-2}}$)}} \\
  & $\bold{\Re}{\mbox{\bfseries{e}}}$ & $\bold{\Im}{\mbox{\bfseries{m}}}$ & $\bold{\Re}{\mbox{\bfseries{e}}}$ & $\bold{\Im}{\mbox{\bfseries{m}}}$ & $\bold{\Re}{\mbox{\bfseries{e}}}$ & $\bold{\Im}{\mbox{\bfseries{m}}}$ & $\bold{\Re}{\mbox{\bfseries{e}}}$ & $\bold{\Im}{\mbox{\bfseries{m}}}$ \\ 
%\hline%[0.7pt]
\midrule[0.7pt]
1 & \phantom{-}0 & 3.02 & \phantom{-}5.12 &               0 & 0 & \phantom{1}3.18 & \phantom{-}6.38 & 0 \\
2 & \phantom{-}4.41 & 6.43 &           -2.56 & \phantom{-}0.97 & 0 &           13.86 & \phantom{-}1.72 & 0 \\
3 &           -4.41 & 6.43 &           -2.56 &           -0.97 & 0 & \phantom{1}7.62 &           -8.10 & 0 \\ 
\bottomrule[0.95pt]
\end{tabular}
\end{center}
\caption{Coefficients of the partial fraction expansion for Kelly's parameterization~\cite{Kel04}. Note the unit change of the polynomial coefficients as compared to Kelly's polynomial: $a_1 \equiv (\hbar/2M)^2 a_1$, $b_1 \equiv (\hbar/2M)^2 b_1$, $b_2 \equiv (\hbar/2M)^4 b_2$, $b_3 \equiv (\hbar/2M)^6 b_3$, where $M$ is the proton mass.}
\label{Gmpoles}
\end{table}

%
% ------------------------------------------------------------------------------------------------------------------------
%
\section{Truncated moments of a polynomial ratio form factor} \label{App2}

Analytical expressions for truncated integer moments are derived hereafter for the polynomial ratio parameterization of the Fourier transform $\tilde{f}_K({\bold k})$ of Eq.~\ref{ffkel}, within the exponential regularization approach of Eq.~\ref{sect2eq03p}. 

Following the discussion of Sec.~\ref{sec:app}, truncated integer moments are defined for the cut-off $Q$ by Eq.~\ref{eq:rmQ} and  Eq.~\ref{Rinte}. The integral is performed before taking the $\epsilon$-limit and takes the generic form
\ba
{\mathcal R}_{2p}(Q,\epsilon) & = & \epsilon \, u_{2p}(Q,\epsilon) + \epsilon \, \sum_{i=1}^3 {_{i}v_{2p}}(\epsilon) \, {\rm Arctan}\left(\frac{Q}{\vert k_i \vert}\right) \nonumber \\
& + & w_{2p}(\epsilon) \, {\rm Arctan}\left(\frac{Q}{\epsilon}\right) \label{evefk} \\
{\mathcal R}_{2p+1}(Q,\epsilon) & = & u_{2p+1}(Q,\epsilon) + \sum_{i=1}^3 {_{i}v_{2p+1}(\epsilon)} \, {\rm Arctan}\left(\frac{Q}{\vert k_i \vert}\right) \nonumber \\
& + & \epsilon \, w_{2p+1}(\epsilon) \, {\rm Arctan} \left(\frac{Q}{\epsilon}\right) \label{oddfk} \, .
\ea
for even and odd truncated moments. Similarly to the dipole parameterization, the $u_j$'s, $_iv_j$'s, and $w_j$'s coefficients accept finite limits when $\epsilon \to 0$. The $u_j$'s are the only coefficients depending on the cut-off, and they vanish for infinite $Q$. The full expression of these functions is too cumbersome to be reported here, but gets simplified when $\epsilon$ tends to zero. \newline 
The $\epsilon$-dependence in Eq.~\ref{evefk} and Eq.~\ref{oddfk} distinguishes the $Q$-saturation behaviour. In the $\epsilon \to 0^+$ limit, the even truncated moments become      
\be
( r^{2p}, f_K ) =  (2p+1)! \, w_{2p}(0^+) 
\ee
independent of $Q$, while the odd truncated moments write 
\ba
& & ( r^{2p+1}, f_K ) = \frac{2}{\pi} \, (2p+2)! \, \times \\
& & \left[ u_{2p+1}(Q,0^+) + \sum_{i=1}^3 {_{i}v_{2p+1}(0^+)} \, {\rm Arctan}\left(\frac{Q}{\vert k_i \vert}\right) \right] \nonumber
\ea
still depending on the cut-off. For instance, the first even moments can be expressed as
\ba
( r^{0}, f_K ) & = & 1 \\ 
( r^{2}, f_K ) & = & 3! \, \left( b_1 - a_1 \right) \\ 
( r^{4}, f_K ) & = & 5! \, \left( b_1^2- a_1 b_1 - b_2 \right) \, 
\ea
and the recurrence relation 
\ba
& & ( r^{2p}, f_K ) = (2p+1)! \, \times \\
& & \left[ b_1\frac{( r^{2p-2}, f_K )}{(2p-1)!} - b_2 \frac{( r^{2p-4}, f_K )}{(2p-3)!} 
+ b_3 \frac{( r^{2p-6}, f_K )}{(2p-5)!} \right] \, , \nonumber
\label{eq:recurrence}
\ea
with $p>2$, provides all the higher orders. The integrals corresponding to the first odd moments write
\ba
{ \cal R}_{1}(Q,0^+) &=& \frac{1}{Q} - 2i \, \frac{A_1}{k_1^{3}} {\rm Arctan} \left(\frac{Q}{|k_1|}\right) \label{eq:r1} \\
& - & 2i \, \frac{A_2}{k_2^3} {\rm Arctan} \left(\frac{Q}{|k_2|}\right) - 2i \,  \frac{A_3}{k_3^3} {\rm Arctan} \left(\frac{Q}{|k_3|}\right) \nonumber \\
{ \cal R}_{3}(Q,0^+)  &=& \frac{b_1-a_1}{Q} - \frac{1}{3Q^3} 
\label{eq:r3} \\
&+& 2i \, \frac{A_1}{k_1^{5}} {\rm Arctan} \left(\frac{Q}{|k_1|}\right) 
 +  2i \, \frac{A_2}{k_2^5} {\rm Arctan} \left(\frac{Q}{|k_2|}\right)\nonumber \\ 
 &+& 2i \, \frac{A_3}{k_3^5} {\rm Arctan} \left(\frac{Q}{|k_3|}\right) \nonumber \\
{ \cal R}_{5}(Q,0^+) & = & \frac{b_1^2- a_1 b_1 - b_2}{Q} - \frac{b_1-a_1}{3Q^3} + \frac{1}{5Q^5} \label{eq:r5} \\
& - & 2i \, \frac{A_1}{k_1^{7}} {\rm Arctan} \left(\frac{Q}{|k_1|}\right) - 2i \, \frac{A_2}{k_2^7} {\rm Arctan} \left(\frac{Q}{|k_2|}\right) \nonumber \\
& - & 2i \, \frac{A_3}{k_3^7} {\rm Arctan} \left(\frac{Q}{|k_3|}\right) \nonumber \, .
\ea
The specific structure of $\tilde{f}_K({\bold k})$ as a ratio of polynomials of even power of $k$ with no poles on the real $k$-axis, leads either 
to pure imaginary poles or to relationship between $A_i$'s and $k_i$'s. For instance, in addition to the general properties of Eq.~\ref{relaKel} we have for the proton electric form factor (Tab.~\ref{Gmpoles})
\ba
 \vert k_2 \vert = \vert k_3 \vert & \Rightarrow & \vert A_2 \vert = \vert A_3 \vert \\
            k_2 = - \overline{k_3} & \Rightarrow & A_2 = \overline{A_3}
\ea
such that ${\cal R}_{2p+1}(Q,0^+)$ are pure real quantities. In the limit $Q \to \infty$, Eq.~\ref{eq:r1}-\ref{eq:r5} provide 
\ba
( r^{1}, f_K ) & = & -2i \, 2! \left[ \frac{A_1}{k_1^{3}} + \frac{A_2}{k_2^{3}} + \frac{A_3}{k_3^{3}}\right] \\
( r^{3}, f_K ) & = & \phantom{-}2i \, 4! \left[ \frac{A_1}{k_1^{5}} + \frac{A_2}{k_2^{5}} + \frac{A_3}{k_3^{5}}\right] \\
( r^{5}, f_K ) & = & -2i \, 6! \left[ \frac{A_1}{k_1^{7}} + \frac{A_2}{k_2^{7}} + \frac{A_3}{k_3^{7}}\right] \, ,
\ea
and generally
\be
( r^{2p+1}, f_K ) = {(-1)}^{p+1} 2i \, (2p+2)! \, \sum_{i=1}^3 \frac{A_i}{{k_i}^{2p+3}} \, . \label{eq:rlkelly}
\ee

%
% ------------------------------------------------------------------------------------------------------------------------
%
%\end{appendix}





\end{document}